\documentclass[reprint,aps,amsmath,amssymb,tightenlines,longbibliography,floatfix,notitlepage]{revtex4-1}

\usepackage[utf8]{inputenc}
\usepackage{bm}

\usepackage{graphicx}
\usepackage{braket}
\usepackage{color}
\usepackage[hidelinks]{hyperref}

\begin{document}
\title{Broadband optical switch based on an achromatic photonic gauge potential in dynamically modulated waveguides}
\author{Ian A. D. Williamson}
\email{iwill@stanford.edu}
\author{Shanhui Fan}
\email{shanhui@stanford.edu}
\affiliation{Department of Electrical Engineering, Stanford University, Stanford, California 94305, USA}
\date{\today}

\begin{abstract}
  We demonstrate that a photonic gauge potential, which arises from the phase degree of freedom in dynamic refractive index modulation and hence is achormatic, can be used to achieve a broadband optical switch using the configuration of a photonic Aharonov-Bohm interferometer (ABI). The resulting ABI switch has a far larger bandwidth and lower cross talk than the conventional Mach-Zehnder interferometer (MZI). Using coupled mode theory and full-wave numerical modeling, we compare the response of the two interferometers in the presence of nonidealities. Our results indicate the importance of the photonic gauge potential for broadband optical signal processing.
\end{abstract}

\maketitle


\section{Introduction}

It was recently shown that in photonic structures undergoing dynamic refractive index modulation, the phase degree of freedom in the modulating waveform corresponds to a gauge potential for photons. Such a gauge potential has been used to construct photonic Aharonov-Bohm interferometers \cite{fang_photonic_2012,tzuang_non-reciprocal_2014}, optical isolators \cite{lira_electrically_2012} and circulators \cite{sounas_giant_2013,williamson_dual-carrier_2018,shi_nonreciprocal_2018}, and to demonstrate non-trivial topological photonic effects \cite{fang_effective_2013,fang_controlling_2013,yuan_photonic_2016,song_direction-dependent_2019}.

An effect of a gauge potential is to induce a phase shift for photons \cite{lin_light_2014,ozawa_synthetic_2016,lumer_observation_2018,liu_realizing_2018,qin_spectrum_2018}. However, unlike the usual phase associated with the propagation of light over a distance, a phase shift due to the gauge potential as induced by dynamic modulation is achromatic. Exploiting this characteristic, in this article we show that the bandwidth of an optical switch constructed from the photonic version of the Aharonov-Bohm interferometer (ABI) \cite{aharonov_significance_1959}, which utilizes such a gauge potential, can be far broader compared to the standard Mach-Zehnder interferometer (MZI). Our work highlights the importance of the concept of the photonic gauge potential for optical information processing.

\begin{figure}[t]
  \centering
  \includegraphics{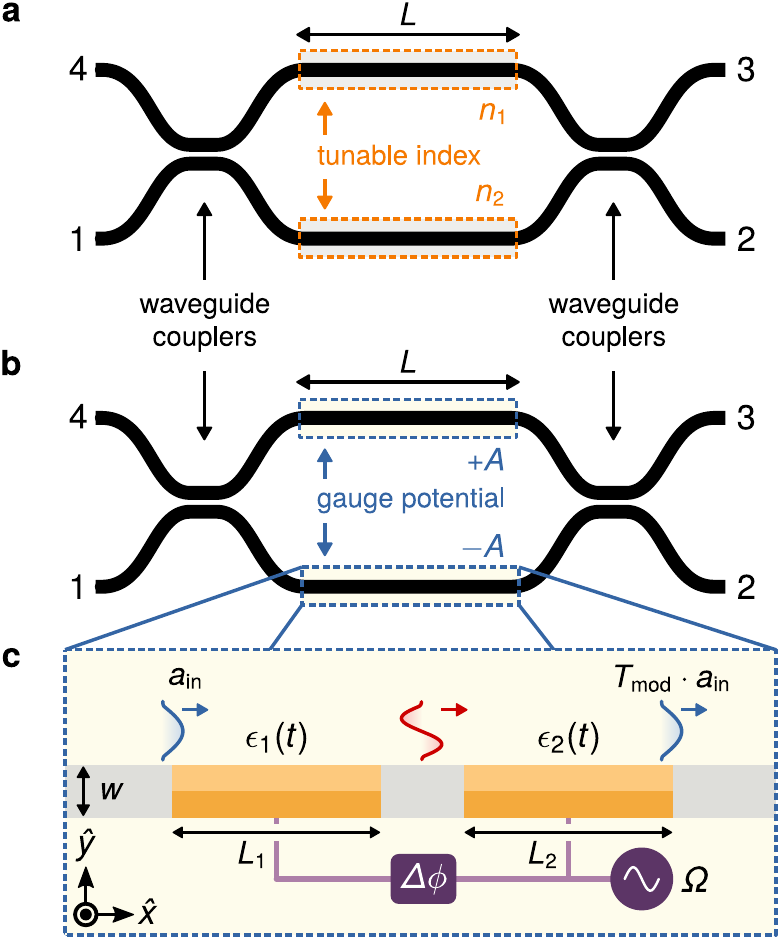}
  \caption{(a) Mach-Zehnder interferometer (MZI) in a push-pull configuration constructed from 50:50 waveguide couplers connected by parallel interferometer arms with tunable refractive indices $n_{1,2}$. (b) Aharonov-Bohm interferometer (ABI) constructed from 50:50 waveguide couplers connected by parallel interferometer arms with photonic gauge potentials $\pm{A}$. (c) Implementation of the gauge potential by a sequence of two dynamically modulated waveguide transitions. The parameter $\Delta{\phi}$ represents the phase shift applied to the modulating waveform of frequency $\Omega$.}
  \label{fig:interferometers}
\end{figure}

\section{Basic concept} \label{sec:concept}

We first review the MZI in the push-pull configuration shown schematically in Fig. \ref{fig:interferometers}(a). The MZI consists of two 50:50 waveguide couplers connected by parallel waveguide arms of length $L$. For an incident optical signal in port one, the transmission to port two and port three are given by
\begin{align}
  T_{21} &= \sin{\left(\frac{\phi_t - \phi_b}{2}\right)}  \label{eq:T21} \\
  T_{31} &= \cos{\left(\frac{\phi_t - \phi_b}{2}\right)}, \label{eq:T31}
\end{align}
where the propagation phases in the top and bottom arms are
\begin{equation}
  \phi_{t,b} = \frac{2\pi{L}}{\lambda_0} n_{t,b}. \label{eq:phi_dispersive}
\end{equation}
Such an MZI can operate as an ideal switch for incident light at a wavelength $\lambda_0$. In the \textit{on} state, the effective indices of the arms are equal $n_t = n_b = n_0$, and port one transmits completely to port three but not to port two, i.e. $T_{21} = 0$ and $T_{31} = 1$. In the \textit{off} state, the effective index of the waveguides are unequal, where $n_t = n_0 + \Delta{n}$ and $n_b = n_0 - \Delta{n}$ for $\Delta{n} = \lambda_0/4L$, and light is instead transmitted from port one to port two but not to port three, i.e. $T_{21} = 1$ and $T_{31} = 0$. However, such a switch is ideal only at a wavelength of $\lambda_0$ and in the vicinity of $\lambda_0$, the contrast between the \textit{on} and \textit{off} states degrades because $\phi_t - \phi_b$ is wavelength dependent.

Like the MZI, the ABI consists of the same 50:50 waveguide couplers and two parallel arms with effective index $n_0$ and length $L$, as shown in Fig. \ref{fig:interferometers}(b). Unlike the MZI, the ABI has gauge potentials with opposite signs, $+A$ and $-A$, in its top and bottom arms as shown schematically in Fig. \ref{fig:interferometers}(c). As a result of these gauge potentials, photons propagating through the two arms experience a phase shift of
\begin{equation}
  \phi_{t,b} = \frac{2\pi{L}}{\lambda_0} n_0 \pm \int{A \cdot dl}. \label{eq:phi_gauge}
\end{equation}
From Eq. \ref{eq:T31}, in the \textit{on} state $\int{A \cdot dl} = 0$ and in the \textit{off} state $\int{A \cdot dl} = \pi/2$. Like the MZI, the system operates as an ideal switch at $\lambda_0$. However, unlike the MZI, as long as $A$ is independent of wavelength, ideal switching in the ABI is preserved for any wavelength. Thus, the use of a photonic gauge potential can in principal create an optical switch with infinite signal bandwidth.

In practice, an important measure of the performance of a switch is the contrast ratio, defined as the ratio between $\vert{T_{31}}\vert^2$ in the \textit{on} and \textit{off} states. Since in the \textit{on} state $T_{31} \approx 1$, the contrast ratio is dominated by deviation of $T_{31}$ from zero in the \textit{off} state. Such deviation as a function of frequency, which primarily results from waveguide group velocity dispersion in the ABI, determines the bandwidth of the switch. In this paper, we will show that in realistic structures, the ABI can have a lower $\vert{T_{31}}\vert^2$ in the \textit{off} state over a broad bandwidth as compared to the MZI. Hence, the ABI can have a larger switching bandwidth.

The minimization of $\vert{T_{31}}\vert^2$ in the \textit{off} state over a broad bandwidth is important in various optical systems. A non-zero $\vert{T_{31}}\vert^2$ in the \textit{off} state results in cross talk which has significant consequences when the interferometer is used as an electrically-reconfigurable routing element \cite{annoni_unscrambling_2017,hughes_training_2018,pai_matrix_2018}. In general, the small signals leaked by the switch in the \textit{off} state generate interference with downstream circuits by limiting the smallest resolvable signal. For example, in sensing and detection systems such as photonic radar \cite{urick_fundamentals_2015,ghelfi_fully_2014}, small amounts of interference can severely limit detection sensitivities by overwhelming the extremely faint signals which are of interest. In optical communication and information processing systems, the associated reduction in dynamic range from such interference limits the maximum amount of information which can be transmitted, or processed, per unit time. Thus, the unique spectral response of the ABI may be of interest to many real-world optical systems where signals have substantial bandwidths, as compared to monochromatic sources which have not been broadened, e.g. by high-speed modulation.

\begin{figure}[t]
  \centering
  \includegraphics{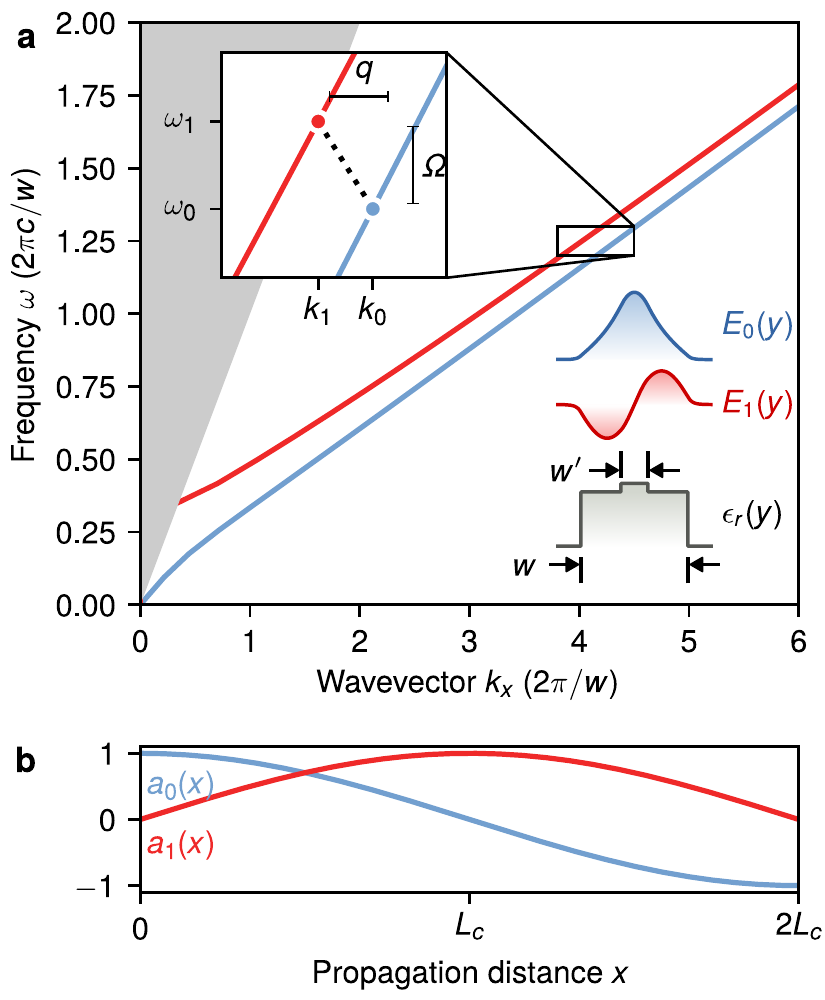}
  \caption{(a) Dispersion relation in normalized units and transverse field distributions, $E_{0,1}{\left(y\right)}$ (shown lower right), of the two lowest order waveguide modes for the waveguide with the transverse permittivity profile, $\epsilon_{r}{\left(y\right)}$ (also shown lower right). The inner core of the waveguide has a width $w^\prime$ and relative permittivity $\epsilon_r = 12.88$ while the outer core has a width $w$ and relative permittivity $\epsilon_r = 12.25$. The waveguide is spatially invariant in the $z$-direction and propagation occurs in the $x$-direction. (b) Spatial evolution of the mode amplitudes, $a_0{(x)}$ and $a_1{(x)}$ as they propagate along a section of the modulated waveguide for an input in the zero-order mode, $a_0{(x=0)} = 1$ and $a_1{(x=0)} =0 $.}
  \label{fig:overview}
\end{figure}

\section{Photonic gauge potential}

To construct an ABI, we use photonic transitions in a dynamically modulated waveguide \cite{winn_interband_1999,dong_inducing_2008,yu_complete_2009} to achieve an achormatic gauge potential, as shown in Fig. \ref{fig:interferometers}(c). The general requirement for such a photonic transition is a waveguide supporting two modes. For example, consider a waveguide with the dispersion relation shown in Fig. \ref{fig:overview}(a), which supports an even and an odd mode and is invariant in the $z$-direction. In order to create a gauge potential for photons, the permittivity of two sections of waveguide are modulated with a time-dependence
\begin{equation}
  \frac{\epsilon_{1,2}{\left(t\right)}}{\epsilon_0} = \epsilon_r + \Delta{\epsilon_r}{\left(y\right)}\sin{\left(\Omega{t} - qx +\phi_{1,2}\right)}, \label{eq:epsilon}
\end{equation}
where the modulation frequency satisfies
\begin{equation}
  \Omega = \omega_1{\left(k_1\right)}-\omega_0{\left(k_0\right)},
  \label{eq:Omega}
\end{equation}
and the modulation wavevector satisfies
\begin{equation}
  q = k_1-k_0.
  \label{eq:q}
\end{equation}
We assume weak modulation, $\Delta{\epsilon_r}/\epsilon_r \ll 1$, which is typical of realistic electro-optical modulators and where a perturbative treatment of the modulation's effect on the waveguide modes is valid.

As light propagates along the $x$-direction in the first modulated waveguide section, as shown in Fig. \ref{fig:interferometers}(c), energy is parametrically converted from the zero-order mode to the first-order mode. The field is uniform in the $z$-direction and, assuming the modulation starts at $x=0$, for $x>0$ the field in the waveguide has the form \cite{yu_complete_2009,fang_photonic_2012}
\begin{multline}
  E_{0}{\left(x,y\right)} = a_{0}{\left(x\right)}E_{0}{\left(y\right)} e^{-j{k_0}{x}} e^{j\omega_{0}{t}} \\
  + a_{1}{\left(x\right)}E_{1}{\left(y\right)} e^{-j{k_1}{x}} e^{j\omega_{1}{t}},
\end{multline}
where $E_{0,1}{\left(y\right)}$ are the transverse profile of the $z$-component of the modal electric field (shown in the lower right of Fig. \ref{fig:overview}(a)). The modal amplitudes evolve along the direction of propagation as
\begin{align}
  a_0{\left(x\right)} &= \cos\left(Cx\right)                    \label{eq:a0} \\
  a_1{\left(x\right)} &= je^{j\phi_1}\sin\left(Cx\right) \label{eq:a1},
\end{align}
where $C$, the coupling strength, is
\begin{equation}
  C = \frac{\omega^2}{4k_x} \int_{0}^{w}{ \Delta{\epsilon_r}{\left(y\right)} E_{z0}{\left(y\right)} E_{z1}{\left(y\right)} dy}. \label{eq:C}
\end{equation}
In order for the two modes to couple, i.e. in order to achieve a non-zero $C$,  the transverse profile of the modulation $\Delta{\epsilon_r}{\left(y\right)}$ must be nonuniform. In this work we always assume an asymmetric distribution with constant magnitude, $\vert \Delta{\epsilon_r} \vert$ across the waveguide. As shown in Fig. \ref{fig:overview}(b), after propagating for a distance of
\begin{equation}
  x = L_c = \frac{\pi}{2}\frac{1}{C},
  \label{eq:Lc}
\end{equation}
an incident wave in the zero-order mode is completely converted to the first-order mode and accumulates a phase of $\phi_1$, as indicated by Eq. \ref{eq:a1}. Similarly, an incident wave in the first-order mode is completely converted to the zero-order mode over the same distance, but the accumulated phase $-\phi_1$ is opposite in sign.

As shown in Fig. \ref{fig:interferometers}(c), two modulators with lengths $L_1 = L_2 = L_c$ are placed in each arm of the ABI. As a result, output light after propagating through the two modulated sections has the same frequency and spatial modal profile as the input light, but acquires a phase of 
\begin{equation}
  \gamma = \pi + \Delta{\phi}, \label{eq:delta_phi}
\end{equation}
in addition to the usual propagation phase. Following Ref. \citenum{fang_photonic_2012}, the relative phase defined by Eq. \ref{eq:delta_phi} is $\Delta{\phi} = \phi_1 - \phi_2$, and can be interpreted as a gauge potential for light. Thus, the structure shown in Fig. \ref{fig:interferometers}(e) provides a physical implementation of the ABI interferometer as shown in Fig. \ref{fig:interferometers}(c). 

\section{Bandwidth consideration}

\begin{figure}
  \centering
  \includegraphics{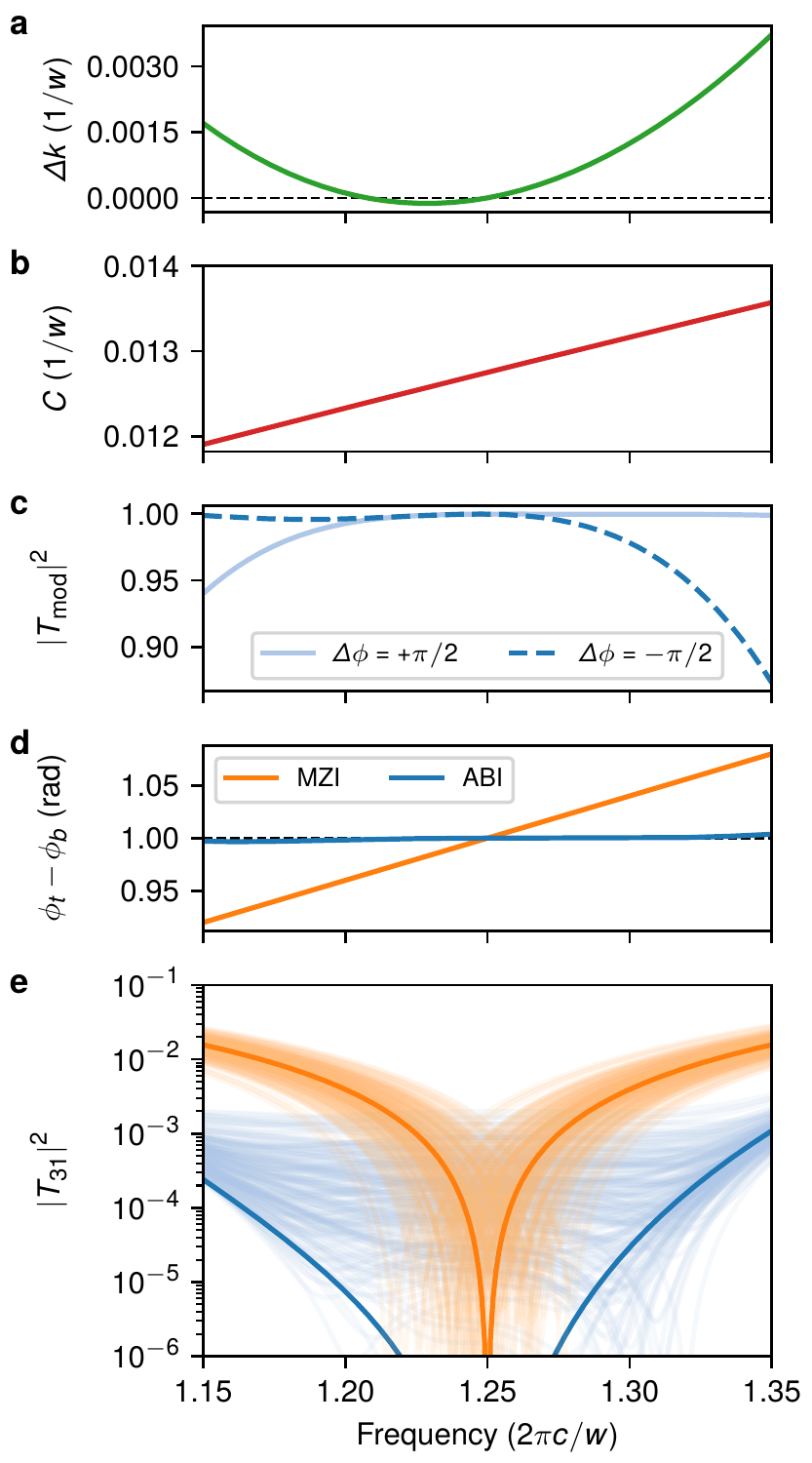}
  \caption{(a) Phase matching $\Delta{k}$ and (b) coupling strength $C$ in units of $1/w$ (inverse waveguide width) as a function of the normalized input frequency for the waveguide shown in Fig. \ref{fig:overview}(a). The input is the zero-order waveguide mode. (c) Transfer function through the sequence of two modulators $\vert{T_{\text{mod}}}\vert^2$ shown in Fig. \ref{fig:interferometers}(c), for gauge potentials consisting of $\Delta{\phi} = \pm \pi/2$. (d) Relative phase between bottom and top arms of the ABI (blue), MZI (orange). The targeted phase of $\Delta{\phi}=\pi$ is indicated by the dashed black line. (e) Transfer function $\vert{T_{31}}\vert^2$ between port 1 and port 3 of the ABI (blue) and MZI (orange). Dark colored lines correspond to the response of an ideal interferometer and lighter colored lines correspond to responses with 200 randomly distributed imbalanced factors of $(1 - \vert\mathcal{N}_a\vert) e^{j \mathcal{N}_\phi}$ applied to the top arm of the interferometer. $\mathcal{N}_a$ corresponds to a normal distribution with mean of zero and standard deviation of $0.01$ and $\mathcal{N}_\phi$ corresponds to a normal distribution with mean of zero and standard deviation of $0.01\pi$.}
  \label{fig:fig_cmt_interferometer}
\end{figure}

The discussion in the previous section indicates that, in principle, one should be able to achieve an achromatic gauge potential based on photonic transitions in a waveguide. In practice, to actually achieve such an achromatic gauge potential, one needs to take into account two aspects which may limit the bandwidth: (1) The phase matching condition of Eqs. \ref{eq:Omega}-\ref{eq:q}, which can be exactly satisfied at one frequency, may not be satisfied over a broad range of frequencies due the group velocity mismatch between the two modes, and the group velocity dispersion in each of the bands. (2) The coupling constant $C$ given by Eq. \ref{eq:C} may also vary as a function of frequency. Consequently the length of the modulators, designed according to Eq. \ref{eq:Lc} and hence optimal for one frequency, may not be optimal for other frequencies. 

Based on the discussion in Sec. \ref{sec:concept}, we now focus on the effect of such non idealities on $\vert{T_{31}}\vert^2$ in the \textit{off} state. We consider the effect of phase matching first. Suppose the phase matching condition of Eq. \ref{eq:Omega} is satisfied at $\omega_0$, for a modulation frequency $\Omega$. In the vicinity of the frequency $\omega_0$, the wavevector mismatch in the photonic transition is given by
\begin{multline}
  \Delta{k}{\left(\omega\right)} = \left[ {{v_{g0}}^{-1}{\left(\omega_0\right)}} - {{v_{g1}}^{-1}{\left(\omega_0+\Omega\right)}} \right] (\omega-\omega_0) + \\
      \frac{1}{2} \left[ {{D_{0}}^{-1}{\left(\omega_0\right)}}^{-1} - {{D_{1}}^{-1}{\left(\omega_0+\Omega\right)}} \right] (\omega-\omega_0)^2,
      \label{eqn:deltak}
\end{multline}
where $v_{g0,1}$ are the group velocities of the modes, $D_{0,1}$ are the group velocity dispersion of the modes, and $\omega$ is the input frequency. Therefore, to reduce the phase mismatch it is important to have a mechanism for controlling the group velocity of the two bands independently. 

For the purpose of achieving independent control of the group velocity of the two bands, we consider the geometry shown in the lower right corner of Fig. \ref{fig:overview}(a), which consists of a dielectric region surrounded by air. The dielectric region has an inner core with $\epsilon_r = 12.88$ and width $w^\prime$ and an outer core of slightly lower permittivity $\epsilon_r = 12.25$ and width $w$. This waveguide design is chosen because the group velocity of the zero-order mode with even symmetry, which has higher field concentration at the waveguide center, can be tuned independently from that of the first-order mode with odd symmetry by increasing or decreasing $w^\prime$. We assume a normalized modulation frequency $\Omega = 3.49\times10^{-2}\ (2{\pi}c/w)$ and a direct transition with $q = 0$, both of which will be used in the subsequent numerical demonstrations. For $w^\prime = 0.26w$, Fig. \ref{fig:fig_cmt_interferometer}(a) shows the computed $\Delta{k}{\left(\omega\right)}$ in the vicinity of $\omega_0 = 1.25\ (2{\pi}c/w)$. $\Delta k(\omega)$ indeed has a parabolic shape, and moreover is relatively small around $\omega_0$. Thus, with proper design one can indeed achieve a small phase mismatch over a substantial bandwidth. 

We now consider the effect of dispersion in the coupling strength, $C$, as defined in Eq. \ref{eq:C}. The dispersion in $C$ arises because the transverse mode profiles vary as a function of frequency. For the system considered here, Fig. \ref{fig:fig_cmt_interferometer}(b) indicates that in the frequency range where $\Delta{k}{(\omega)}$ is small, $C$ varies by approximately 10\%. Thus, the modulator length selected to provide complete conversion between the two waveguide modes for an input at $\omega_0 = 1.25\ (2{\pi}c/w)$ will not provide complete conversion at higher and lower frequencies. 

This dispersion in $C$ affects the transfer function $\vert{T_{\text{mod}}}\vert^2$ through the sequence of two modulators as shown in Fig. \ref{fig:fig_cmt_interferometer}(c). Moreover, at $\omega \ne \omega_0$, $\vert{T_{\text{mod}}}\vert^2$ depends on the choice of gauge potential, $\Delta{\phi}$. The change in $\vert{T_{\text{mod}}}\vert^2$ with $\Delta{\phi}$ effectively yields an imbalanced insertion loss between the two arms. We emphasize that this effective insertion loss is not dependent on material absorption, but instead results from incomplete conversion between the modes. Any energy in the first-order mode after light passes through both modulators is treated as loss because we focus on the interference process in the zero-order mode. In the full-wave implementation of the ABI we show later, any remaining energy in the first order mode is scattered by waveguide tapering at the ends of the interferometer arms.

Unlike the amplitude response of the modulators, which exhibit dispersion, as shown in Fig. \ref{fig:fig_cmt_interferometer}(c), the relative phase between the top and bottom arms remains very close to $\pi$ over the entire frequency range, as shown in Fig. \ref{fig:fig_cmt_interferometer}(d). We note that to achieve such a near-achromatic phase shift, the ABI must be operated in a push-pull configuration, where both arms include modulators. This allows any effect of the permittivity modulation on the static waveguide propagation phase, corresponding to the first term in Eq. \ref{eq:phi_gauge}, to be applied equally to both arms. In contrast, the phase difference between the two arms of the MZI, indicated by the orange line in Fig. \ref{fig:fig_cmt_interferometer}(d), exhibits strong dispersion with a linear frequency dependence from Eq. (\ref{eq:phi_dispersive}) and a total variation in the phase determined by the relative bandwidth, which is approximately 15\%. The relative bandwidth is defined for a frequency range as its width divided by its center frequency.

The dark orange and dark blue curves shown in Fig. \ref{fig:fig_cmt_interferometer}(e) correspond to the the transfer function $\vert{T_{31}}\vert^2$ from the lower left port to the top right port in the MZI and ABI, respectively. In the MZI we observe a very narrow dip in the transfer function at $\omega_0 = 1.25\ (2{\pi}c/w)$, with a significant increase in $\vert{T_{31}}\vert^2$ at lower and higher frequencies. In the MZI, the amplitude responses of the two arms are balanced over the entire spectrum. At $\omega = \omega_0$, the phase difference between the two arms is exactly $\pi$, leading to an exact zero in $\vert{T_{31}}\vert^2$. Thus, the increase in $\vert{T_{31}}\vert^2$ arises because, away from $\omega_0$, the phase difference between the arms is not $\pi$. In the ABI we also observe a dip, but with a much larger bandwidth where $\vert{T_{31}}\vert^2$ is below $10^{-6}$. In the case of the ABI, at the frequency $\omega_0$, the two arms have a balanced amplitude response, and the phase difference is exactly $\pi$, leading to a zero in $\vert{T_{31}}\vert^2$. Away from $\omega_0$, the phase difference remains at approximately $\pi$ and the increase in $\vert{T_{31}}\vert^2$ instead arises from the different amplitude response of the two arms for two different choices of gauge potential, as shown in Fig. \ref{fig:fig_cmt_interferometer}(c). Across the entire bandwidth considered here, $\vert{T_{31}}\vert$ is lower in the ABI than in the MZI, even in the presence of the unbalanced signal conversion through the modulators shown in Fig. \ref{fig:fig_cmt_interferometer}(c).

In the analysis above, for both the MZI and the ABI, we have assumed an idealized scenario, where there exists a frequency $\omega_0$ at which the amplitude responses of the two arms are perfectly balanced, and the phase difference is exactly $\pi$. In practice, due to random fabrication imperfections, there is usually some level of imbalance between the amplitude response of each arm, and moreover there may be a background phase difference in addition to those considered above.

To illustrate the effect of such nonidealities, the light orange and light blue curves shown in Fig. \ref{fig:fig_cmt_interferometer}(e) correspond to the $\vert{T_{31}}\vert^2$ transfer functions for the MZI and ABI, respectively, with 200 randomly unbalanced phase and amplitude factors. The disorder is defined mathematically as a factor of $(1 - \vert\mathcal{N}_a\vert) e^{j \mathcal{N}_\phi}$, which is multiplied by the transfer function of one interferometer arm. The variable $\mathcal{N}_a$ represents a normal distribution with a mean of zero and a standard deviation of $0.01$ and $\mathcal{N}_\phi$ represents a normal distribution with a mean of zero and a standard deviation of $0.01\pi$. In the presence of such disorder, both the ABI and the MZI are no longer guaranteed to exhibit an \textit{exact} zero in $\vert{T_{31}}\vert^2$. We observe for the MZI that all spectra continue to exhibit a pronounced narrow dip. Due to the linear dispersion in the phase difference between its arms, the additional background phase simply shifts the frequency where the phase difference in the MZI is $\pi$, while the unbalanced amplitude response of the arms increases its minimum value of $\vert{T_{31}}\vert^2$.

In contrast, there is not always a single pronounced dip in the spectrum of $\vert{T_{31}}\vert^2$ for the ABI in the presence of disorder. For some of the unbalanced responses shown in Fig. \ref{fig:fig_cmt_interferometer}(e), two separate but much shallower minima in $\vert{T_{31}}\vert^2$ are obverved. The reason for this is that the relative phase shown in Fig. \ref{fig:fig_cmt_interferometer}(d) actually has an inflection point, although this feature is not visible with the negligible slope in the relative phase. Therefore, certain levels of phase imbalance can actually cause the relative phase to cross $\pi$ at multiple frequencies in the spectrum. Moreover, certain levels of amplitude imbalance can counteract the imbalance from incomplete modulator conversion, as is shown in Fig. \ref{fig:fig_cmt_interferometer}(c). However, because many of the nonideal ABI responses shown in Fig. \ref{fig:fig_cmt_interferometer}(e) have a small imbalance in both amplitude \textit{and} phase at \textit{all} frequencies, they exhibit a lineshape in $\vert{T_{31}}\vert^2$ which is very different than the ideal case. Nevertheless, many of the nonideal ABI responses retain their broad bandwidth and relatively flat shapesa as compared with those of the MZI. Except for the  narrow range of frequencies near the MZI's minimum in $\vert{T_{31}}\vert^2$, the value of $\vert{T_{31}}\vert^2$ in the ABI tends to be lower over the entire bandwidth of Fig. \ref{fig:fig_cmt_interferometer}(e). This contrast is especially strong near the edges of the spectrum.

Such a flat response in $\vert{T_{31}}\vert^2$ may be potentially advantageous for broadband signal routing because it eliminates the strong dispersion in the cross talk, thus overcoming the tradeoff between maximum signal bandwidth and the amount of interference. In our analysis so far, we have assumed dispersion-free waveguide couplers with an ideal 50:50 splitting ratio. Although, in practical waveguide couplers one should expect dispersion in the splitting ratio, we note that any deviation from 50:50 splitting does not increase $T_{31}$ whenever $\phi_t - \phi_b \equiv \pi$.

\section{Full-wave numerical simulation}

\begin{figure}
  \centering
  \includegraphics{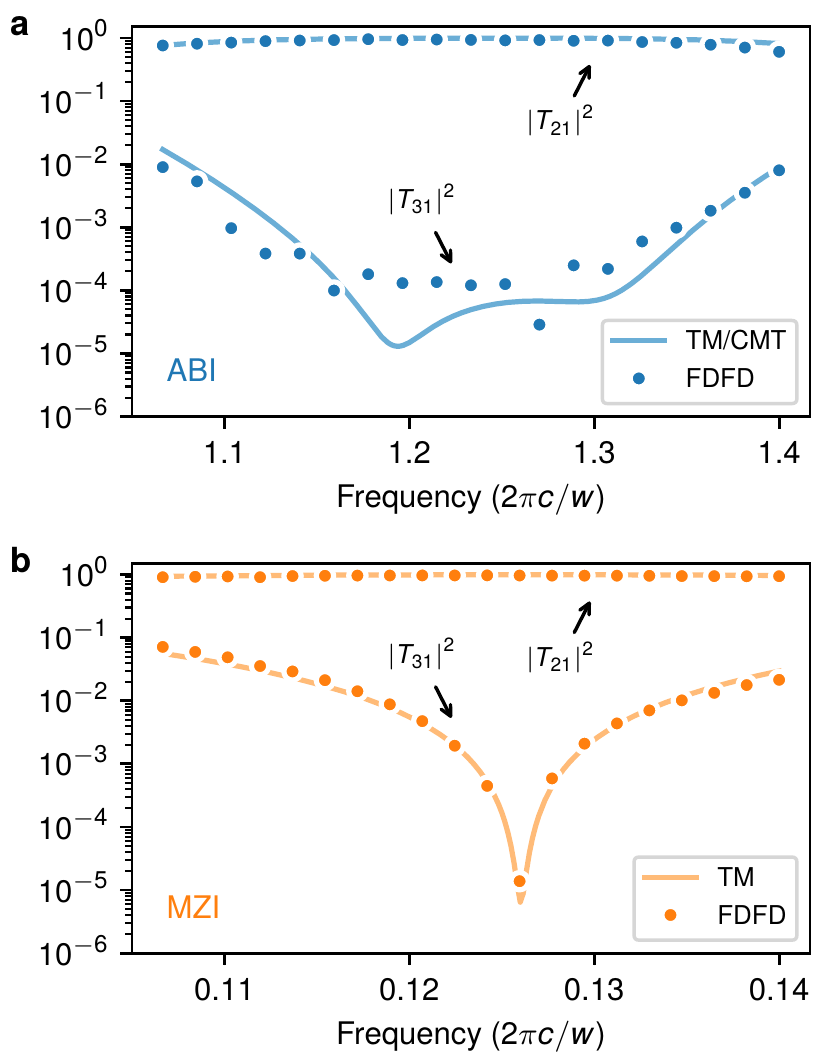}
  \caption{Transfer function from port 1 to port 2, $\left\vert T_{21} \right\vert^2$, and from port 1 to port 3, $\left\vert T_{31} \right\vert^2$, in the \textit{off} state, for (a) the ABI and (b) the MZI. The frequency ranges for panel a and b, which are displayed in units normalized to the waveguide width, are different because the ABI uses a 10$\times$ larger waveguide. However, the relative bandwidths shown in panel a and b are the same. The relative bandwidth is defined for a frequency range as its width divided by its center frequency. The dots were computed using full wave finite difference frequency domain (FDFD) simulations and the solid lines were computed using transfer matrix (TM) and coupled mode theory (CMT) modeling. The TM and CMT results include an unbalanced phase and amplitude factor of $0.995e^{-j0.005\pi}$ multiplied by the transfer function of the top arm.}
  \label{fig:fig_cmt_fit}
\end{figure}

\begin{figure*}
  \centering
  \includegraphics{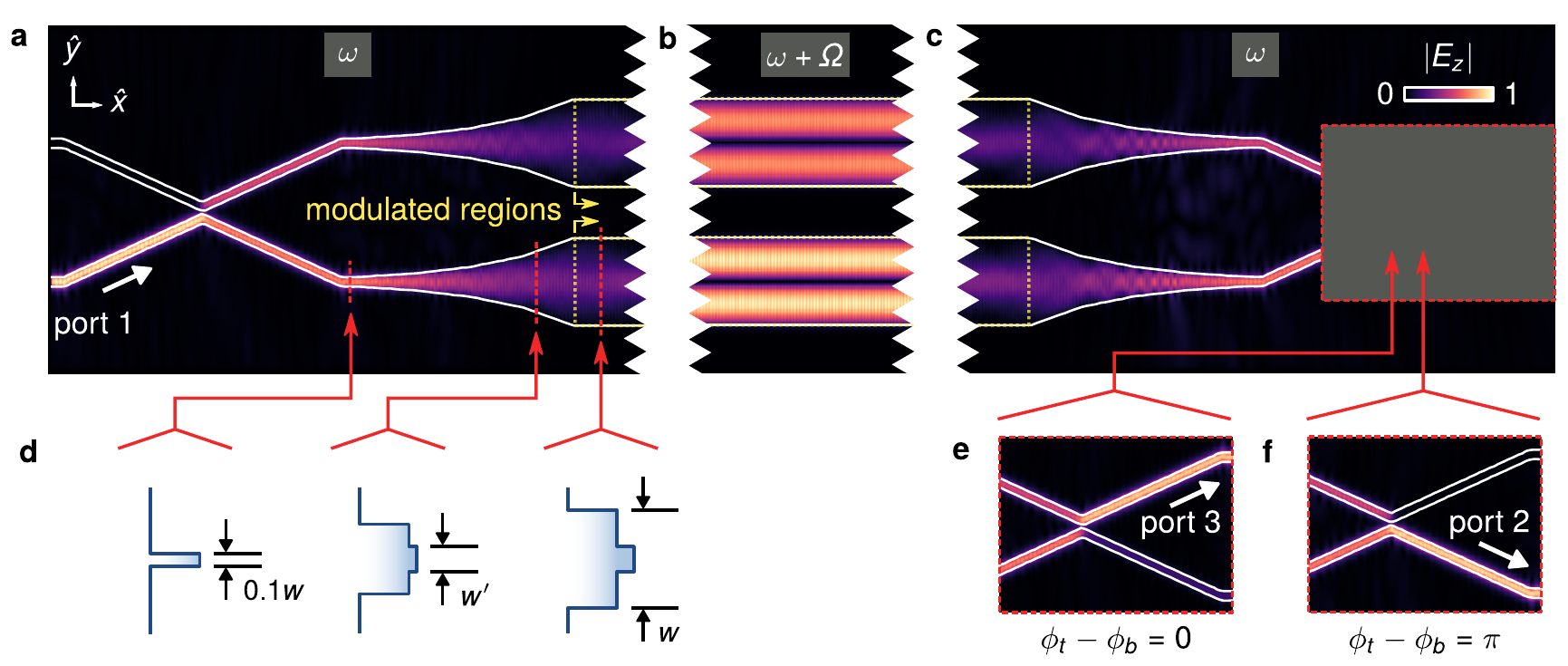}
  \caption{(a-c) Electric field distribution (magnitude of the $z$-component) in the ABI for a signal incident from port 1. Panels a and c show the field at the input and output frequency, $\omega$, while panel b shows the field at the up-converted frequency, $\omega+\Omega$, in the middle region of the interferometer. The edges of the waveguides are indicated by white lines and the modulated waveguide sections are indicated by the yellow dotted lines. (d) Cross section of the waveguide permittivity as it is tapered from a narrow waveguide of width $0.1w$ to the wider modulated waveguide of width $w$ with a permittivity bump of width $w^\prime$. The wider modulated waveguide configuration is identical to the one shown in the inset of Fig. \ref{fig:overview}(a). (e,f) Electric field distribution in the output coupler of the ABI for $\phi_t - \phi_b = 0$, with the output signal routed through port 3, and $\phi_t - \phi_b = \pi$, with the output signal routed through port 4. The $x$-axes of all electric field distributions have been scaled by a factor of $1/3$ with respect to their $y$-axes.}
  \label{fig:fields}  
\end{figure*}

We validate the theoretical considerations above using a full-wave frequency domain solution to Maxwell's equations, fully accounting for coupling between the relevant frequency components via the modulation \cite{shi_multi-frequency_2016, schenk_pardiso_2001}. We consider a two-dimensional computational domain discretized on a Yee lattice surrounded by perfectly matched layer (PML) boundary conditions \cite{shin_choice_2012}. For the modulated waveguides in the ABI, we use the same design shown in the lower right of Fig. \ref{fig:overview}(a), but in the waveguide couplers and at the ports we use a narrower waveguide of width $w/10$. This narrower waveguide consists of only a single dielectric material ($\epsilon_r = 12.25$), and allows for a shorter coupling length in the beam splitters due to an increased overlap of the evanescent fields. At the beginning and end of the interferometer arms, the narrower waveguide is tapered to the width $w$ of the modulated sections over a distance of $L_{\text{taper}} = 8w$. Beginning at a distance of $0.75L_{\text{taper}}$ into the tapered waveguide segment, the permittivity bump of width $w^\prime$ is linearly ramped up from 12.25 to 12.88 for the remaining distance of the tapered waveguide, $0.25L_{\text{taper}}$, to match with the modulated waveguide. The permittivity bump is similarly ramped down over a distance of $0.25L_{\text{taper}}$ on the opposite side of the interferometer. Such a continuous linearly increasing permittivity is used for convenience in our numerical simulations and, in practice, other mode-matching techniques can be used between the waveguides. An additional benefit of the waveguide tapering is that the first-order mode, with odd symmetry, is unable to couple into the narrower single-mode waveguide, preventing further interference from residual energy in the first-order mode.

Fig. \ref{fig:fig_cmt_fit}(a) shows the transfer functions from port 1 to port 2 and port 3 in the \textit{off} state for the ABI, which were computed by numerically integrating the flux over the input and output waveguides in the full wave simulation. We observe that $\vert{T_{31}}\vert^2$ remains relatively uniform, around a value of approximately $10^{-4}$, over approximately a 10\% relative bandwidth at the center of the spectrum. The numerical results agree well with the coupled mode theory (CMT) and transfer matrix (TM) model as discussed in the previous section, which includes a factor of amplitude and phase imbalance $0.995e^{-j0.005\pi}$ in the transmission of the top arm. The numerical results indicate that the phase shift from the photonic gauge potential is indeed nearly achromatic, but the simulated structure has a small imbalance between the two arms, possibly originating from very small reflections off of the waveguide couplers and tapers. Very small reflections from the abrupt boundary of the modulation itself could also contribute to the imbalance observed in the full wave calculation. We note that the waveguide couplers used in the simulation does not have a 50:50 splitting ratio over the entire spectrum shown in Fig. \ref{fig:fig_cmt_fit}. Such dispersion in the waveguide couplers, however, does not significantly influence the performance of the switch because the phase difference between the arms is very close to $\pi$ over a broad bandwidth. We also observe from Fig. \ref{fig:fig_cmt_fit}(a) that $\vert{T_{21}}\vert^2$ in the ABI remains relatively flat and close to unity around the center of the spectrum. Towards the edges of the spectrum, $\vert{T_{21}}\vert^2$ begins to decrease due to incomplete conversion between the waveguide modes from the modulation, confirming the trend seen in the modulator transfer function, $\vert T_{\text{mod}} \vert$ of Fig. \ref{fig:fig_cmt_interferometer}(c).

In contrast, Fig. \ref{fig:fig_cmt_fit}(b) shows that a full wave simulation of the MZI, with identical waveguide couplers to the ABI, exhibits a response in $\vert{T_{31}}\vert^2$ which varies by several orders of magnitude. This simulation of the MZI uses a narrower waveguide of width $w/10$ throughout, which results in a lower normalized operating frequency which is $10\times$ lower when displayed in normalized units. However, we note that the relative bandwidths shown in Fig. \ref{fig:fig_cmt_fit}(a-b) are equivalent. Like that of the the ABI, the full-wave simulation of the MZI is consistent with a transfer matrix calculation which includes a phase and amplitude imbalance, largely confirming the theoretical analysis of the previous section. In the case of the MZI, the increased permittivity required to satisfy Eq. \ref{eq:phi_dispersive} likely results in small reflections which act to slightly shift the frequency where $\vert{T_{31}}\vert^2$ reaches its minimum. Moreover, we observe that $\vert{T_{21}}\vert^2$ in the MZI remains relatively uniform around the center of the spectrum before decreasing slightly towards the edges. We note that $\vert{T_{21}}\vert^2$ tends to be larger in the MZI than in the ABI near the edges of the spectrum due to the incomplete mode conversion through the modulators used in the ABI, as previously discussed.

\begin{figure}
  \centering
  \includegraphics{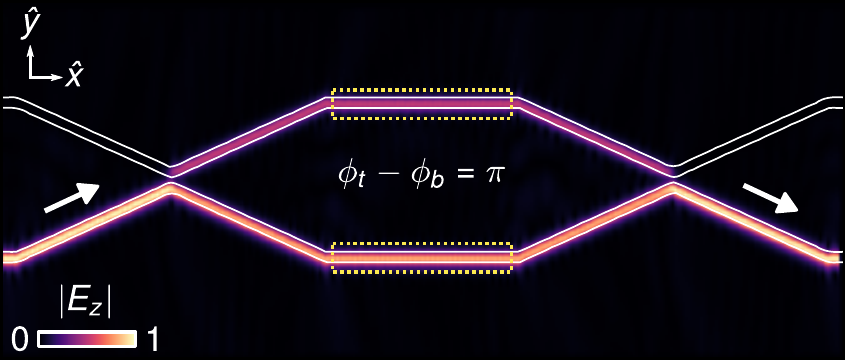}
  \caption{Electric field distribution (magnitude of the $z$-component) in the MZI for a signal incident from port 1. The edges of the waveguides are indicated by white lines and the segment of waveguide with tuned permittivity are indicated by the yellow box. The $x$-axis has been scaled by a factor of $1/3$ with respect to the $y$-axis.}
  \label{fig:fig_fields_mzi}
\end{figure}

Fig. \ref{fig:fields}(a)-(c) show the electric field distribution of the ABI for a signal incident from port 1 at a normalized frequency $\omega = 1.25\ (2{\pi}c/w)$. Fig. \ref{fig:fields}(b) shows the field at the up-converted frequency $\omega+\Omega$ around the halfway point of the interferometer, while Fig. \ref{fig:fields}a and Fig. \ref{fig:fields}(c) show the field around the waveguide couplers at the input and output signal frequency $\omega$. The switching of the optical signal at $\omega$ between the two output ports is shown in Fig. \ref{fig:fields}(e)-(f), where the gauge potentials are tuned from $\phi_t - \phi_b = 0$ to $\phi_t - \phi_b = \pi$. For comparison, the field distribution in the MZI at the frequency with the lowest $\vert{T_{31}}\vert^2$ is shown in Fig. \ref{fig:fig_fields_mzi}. These field patterns confirm that the simulated structures indeed function as optical switches. 

In the ABI demonstrated in Fig. \ref{fig:fields} we have considered a spectrum around a normalized optical frequency of $\omega = 1.25\ (2{\pi}c/w)$, a normalized modulation frequency of $\Omega = 3.49\times10^{-2}\ (2{\pi}c/w)$, a modulation strength of $\Delta{\epsilon_r}/\epsilon_r = 2.45\times10^{-3}$, and length $L_c = 121.86w$. To consider a device operating around $\lambda_0=1550$ nm, we can take $w = 2\ \mu{\text{m}}$. This choice of $w$ results in a modulation frequency which is approximately two orders of magnitude larger than what is capable in state of the art modulators where $\Omega/2\pi\sim$ 10-100 GHz. One approach to using a realistic modulation frequency in the same waveguide design is to instead achieve the gauge potential with an indirect photonic transition where $q \ne 0$. This would make the design similar to the dynamic isolator previously proposed in Ref. \citenum{poulton_design_2012} where a 1 GHz phonon mode played the role of the modulation. In such an indirect transition, the gauge potential would be retained through a global relative modulation phase shift between the modulated waveguide sections, which would each carry an $x$-dependent phase from the momentum. An added constraint from using a lower modulation frequency is the need for a smaller modulation index and larger modulator length, such that $\epsilon_r\Omega/\omega\Delta{\epsilon_r} \gg 1$, to preserve the single-sideband nature of the modulation. 

We emphasize that, in practice, the ABI can leverage standard integrated optical components and material platforms, and is not constrained to those used in our numerical demonstration. For example, multi-mode interferometers and other forms of spatial mode filters can be used to couple the modulated waveguides. Moreover, dispersion engineering could be carried out using the degrees of freedom in three-dimensional waveguide structures, such as the out-of-plane thickness. Three dimensional structures also open up the possibility of coupling between modes with different polarizations (e.g. TM-like and TE-like) if electro-optic tensors with the appropriate symmetry are used. Such modes may be more attractive for dispersion engineering.

\begin{figure}
  \centering
  \includegraphics{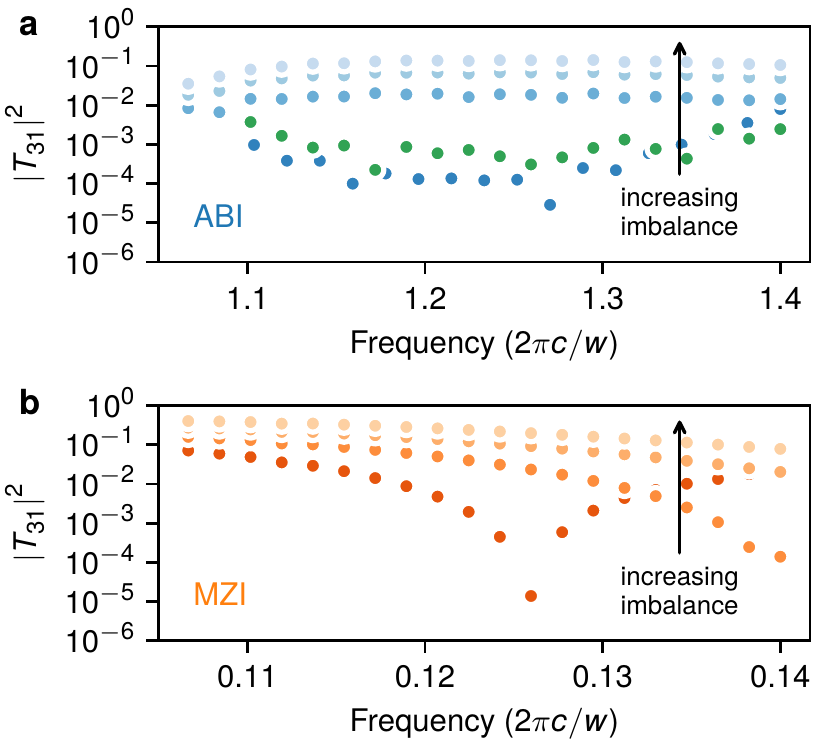}
  \caption{Transfer function from port 1 to port 3, $\left\vert T_{31} \right\vert^2$, in the \textit{off} state, for (a) the ABI and (b) the MZI with an increasing systematic imbalance. The imbalance is defined by the static permitivitty of one interferometer arm being scaled by a factor of 1.005, 1.010, and 1.015, corresponding to successively lighter coloring. The darkest colored dots correspond to the transfer functions from Fig. \ref{fig:fig_cmt_fit}. The dark green dots in (a) correspond to the case of an abrupt transition between the waveguide with uniform permitivitty of $\epsilon_r = 12.25$ and the waveguide with a permitivitty bump of $\epsilon_r = 12.88$, but with a tapering between the smaller and larger width. The transfer functions were computed using full wave finite difference frequency domain simulations.}
  \label{fig:fig_fullwave_disorder}
\end{figure}

In Fig. \ref{fig:fig_fullwave_disorder} we consider the effect of a systematic imbalance between the arms of the interferometers on $\vert{T_{31}}\vert^2$ using full wave simulations. In the ABI, we maintain the modulation frequency and modulation strength at their ideal values (i.e. for achieving $\int{A \cdot dl} = \pi/2$ without the imbalance). Similarly, in the MZI, we maintain the static index shift at its ideal value (i.e. for achieving $\Delta{n} = \lambda/4L$ without the imbalance). The imbalance is defined in the ABI by scaling the permittivity  bump, $\epsilon_r^\prime$ [Fig. \ref{fig:fig_fullwave_disorder}(a)], and in the MZI by scaling the waveguide permittivity , $\epsilon_r$ [Fig. \ref{fig:fig_fullwave_disorder}(b)], by a factor of 1.005, 1.010, and 1.015. These values correspond to the successively lighter colored dots in Fig. \ref{fig:fig_fullwave_disorder}(a) and (b), where the darkest colored dots correspond to $\vert{T_{31}}\vert^2$ from Fig. \ref{fig:fig_cmt_fit}(a) and (b), respectively. In contrast to the random disorder considered in Fig. \ref{fig:fig_cmt_interferometer}(e), we observe that the systematic imbalance causes the switching performance of both interferometers to deteriorate significantly because the ideal biases can no longer provide the required phase shift. This indicates that in practice to maintain a low $\vert{T_{31}}\vert^2$, a calibration routine will be required for selecting the optimal biasing. For the ABI, this will likely require calibration over both the modulation strength and frequency because changes in the waveguide cross section can, in general, affect both the dispersion and the modal overlap. Such a calibration routine, which is beyond the scope of the current work, would also be relevant for imbalances in the geometry of the waveguides, e.g. the widths, which would result in a similar performance degradation. For the ABI, Fig. \ref{fig:fig_fullwave_disorder}(a), we also consider the effect an abrupt transition between the waveguide with a uniform permittivity  of $\epsilon_r = 12.25$ and the waveguide with a permittivity  bump of $\epsilon_r = 12.88$ (but still with a tapering between the smaller and larger width). The spectrum of $\vert{T_{31}}\vert^2$ for this abrupt transition corresponds to the green dots in Fig. \ref{fig:fig_fullwave_disorder}(a), where we observe only a minor deterioration in the switching performance.

\section{Discussion}

In practice, the gauge potential in $\Delta{\phi}$ would originate from a low frequency phase shifter such as a tunable microwave delay line or resonant phase shifter acting on the modulating waveform. From this perspective, the photonic transition samples such a low frequency dispersive phase shift (at the frequency of the continuous wave modulation), and translates it higher in frequency, by many orders of magnitude, into a broadband optical phase shift. In practice the bandwidth of such a low frequency phase shifter is irrelevant so long as it remains continuously tunable from $0-2\pi$ independently of its amplitude response. Unlike the phase shift from a static index shift, the accumulated phase from the gauge potential is not associated with any change in bias amplitude because the peak-to-peak change in the permittivity $\vert\Delta{\epsilon_r}\vert$ remains constant for \textit{all} values of $\Delta{\phi}$.

The switching time between the \textit{on} and \textit{off} states of the ABI is ultimately limited by the reconfiguration time of the radio frequency phase shifter, which is typically on the order of $10-100$ ns \cite{natarajan_fully-integrated_2011-1}. In contrast, all-optical switches based on free carrier dispersion effects can have far smaller switching times, on the order of 10 ps \cite{nozaki_ultralow-energy_2013,moille_integrated_2016,colman_ultrafast_2016}. These switches are attractive for high-speed modulation, but have an optical response which is limited by the cavity linewidth. A narrow Lorentzian or Fano lineshape is useful for achieving a strong contrast between the \textit{on} and \textit{off} states when the cavity is detuned by a strong pump field, but it prevents the system from routing broadband optical signals. Thus, our results indicate that a switch based on the photonic gauge potential may be suitable for platforms which do not require ultrafast response times but benefit from high \textit{on}-\textit{off} contrast ratios and broad signal bandwidths. These include recently proposed reconfigurable optical processors for machine learning, optical communications, and microwave photonics applications with high dynamic range \cite{zhuang_programmable_2015,shen_deep_2017,annoni_unscrambling_2017}.

It is interesting to note that Eq. \ref{eq:delta_phi} can be interpreted as a form of geometric phase shift which is completely controlled by an electrically tunable modulation phase \cite{ranzani_geometric_2014}. This point is important in contrast with geometric phases which arise in adiabatic nonlinear processes, for example in $\chi^{(2)}$ materials \cite{karnieli_fully_2018}. Nonlinear interactions in these devices can also produce arbitrary geometric phase shifts, but instead require tuning the Fourier coefficients of the quasi-phase matched (QPM) gratings. Such phase shifts are therefore fixed at fabrication while the modulation phase of the photonic gauge potential as discussed here can be dynamically reconfigured at will. 

We also note that the dynamics of the ABI and the photonic gauge potential involve only two optical states. In contrast, previous demonstrations of phase shifters based on filtered electro and acousto-optic modulators \cite{ballard_broadband_1978,ehrlich_voltagecontrolled_1988} inherently couple an infinite number of optical states.

\section{Conclusion}
In conclusion, we have shown that a photonic gauge potential constructed from dynamic modulation can be used to create a broadband switch based on a photonic Aharonov-Bohm interferometer. This work highlights the importance of the photonic ABI and the gauge potential for optical signal processing and computing.

\begin{acknowledgments}
  The authors are grateful to Dr. Qian Lin, Dr. Yu Shi, and Dr. Alex Y. Song for many helpful discussions. This work was supported by a U. S. AFOSR MURI Project (Grant N\textsuperscript{\underline{o}} FA9550-17-1-0002).
\end{acknowledgments}

%

\end{document}